\documentclass[lettersize,journal]{IEEEtran}
\usepackage{amsmath,amsfonts}
\usepackage{algorithmic}
\usepackage{algorithm}
\usepackage{array}
\usepackage[caption=false,font=normalsize,labelfont=sf,textfont=sf]{subfig}
\usepackage{textcomp}
\usepackage{stfloats}
\usepackage{url}
\usepackage{verbatim}
\usepackage{graphicx}
\usepackage{cite}
\usepackage{amsmath,amssymb,amsfonts}
\usepackage{algorithmic}
\usepackage{textcomp}
\usepackage{xcolor}
\usepackage{amsmath,graphicx}
\usepackage{multirow}
\usepackage{booktabs}
\usepackage{amssymb}
\usepackage{threeparttable}
\usepackage{subcaption}
\usepackage{amssymb}   
\usepackage{pifont}    

\begin{document}

\title{DeepASMR: LLM-Based Zero-Shot ASMR Speech Generation \\for Anyone of Any Voice}

\author{Leying Zhang,
    Tingxiao Zhou,
    Haiyang Sun,
    Mengxiao Bi,
    and Yanmin Qian, \IEEEmembership{Senior Member, IEEE}
\thanks{Leying Zhang, Tingxiao Zhou, Haiyan Sun and Yanmin Qian are with the Auditory Cognition and Computational Acoustics Lab, School of Computer Science\& MoE Key Laboratory of Artificial Intelligence, AI Institute, Shanghai Jiao Tong University, Shanghai, 200240 P. R. China (e-mail:\{zhangleying, stupid\_computer, sunhaiyang, yanminqian\}@sjtu.edu.cn).

Mengxiao Bi is with VUI Labs, Hangzhou, 310000 P. R. China (e-mail:bimengxiao@vuilabs.cn).

Yanmin Qian is the corresponding author.}}

\markboth{Journal of \LaTeX\ Class Files,~Vol.~14, No.~8, August~2021}%
{Shell \MakeLowercase{\textit{et al.}}: A Sample Article Using IEEEtran.cls for IEEE Journals}


\maketitle

\begin{abstract}
While modern Text-to-Speech (TTS) systems achieve high fidelity for read-style speech, they struggle to generate Autonomous Sensory Meridian Response (ASMR)—a specialized, low-intensity speech style essential for relaxation. The inherent challenges include ASMR's subtle, often unvoiced characteristics and the demand for zero-shot speaker adaptation. In this paper, we introduce DeepASMR, the first framework designed for zero-shot ASMR generation. We demonstrate that a single short snippet of a speaker's ordinary, read-style speech is sufficient to synthesize high-fidelity ASMR in their voice, eliminating the need for whispered training data from the target speaker.  Methodologically, we first identify that discrete speech tokens provide a soft factorization of ASMR style from speaker timbre. Leveraging this insight, we propose a two-stage pipeline incorporating a Large Language Model (LLM) for content-style encoding and a flow-matching acoustic decoder for timbre reconstruction. Furthermore, we contribute DeepASMR-DB, a comprehensive 670-hour English-Chinese multi-speaker ASMR speech corpus, and introduce a novel evaluation protocol integrating objective metrics, human listening tests, LLM-based scoring and unvoiced speech analysis. Extensive experiments confirm that DeepASMR achieves state-of-the-art naturalness and style fidelity in ASMR generation for anyone of any voice, while maintaining competitive performance on normal speech synthesis.
\end{abstract}


\section{Introduction}
\IEEEPARstart{T}{he} landscape of Text-to-Speech (TTS) synthesis has been radically transformed by the emergence of large-scale generative models and massive-scale datasets. Recent architectures leveraging neural codecs, latent diffusion, and flow-matching have achieved unprecedented naturalness, enabling zero-shot voice cloning and high-fidelity speech reconstruction \cite{ju2024naturalspeech, le2023voicebox, chen2024f5, zhang2025covomix2, zhang2025advanced}. Despite these milestones, current TTS research remains largely constrained to neutral and read-style audio. This focus creates a significant gap in the generation of high-affect, non-standard vocalizations, particularly those that rely on unvoiced acoustic profiles, such as Autonomous Sensory Meridian Response (ASMR), where the speaker avoids vibrating their vocal cords to create a quiet, airy sound~\cite{poerio2018more}.

ASMR has evolved from an internet subculture into a global phenomenon, driven by its documented physiological and psychological benefits. Often described as a "tingling" sensation originating in the scalp and radiating down the spine, ASMR is triggered by specific auditory stimuli that induce a state of deep relaxation \cite{sakurai2023brain}. ASMR speech fundamentally differs from normal speech in several key aspects. First, ASMR creators use extremely soft, breathy tones rather than a full speaking voice~\cite{ito2005analysis}. Second, ASMR speech includes both voiced sounds (with vocal-fold vibration) and unvoiced sounds (without vibration), whereas ordinary speech is dominated by voiced sounds~\cite{poerio2018more}. Third, compared to normal speech, whispered vowels are longer in duration and have higher formant frequencies, though the extent of these differences varies by vowel and speaker gender~\cite{houle2020acoustic}.

ASMR has taken the world by storm because of its distinctive physiological benefits: soft whispers and breathy sounds prompt the brain to release endorphins and serotonin while lowering cortisol~\cite{richard2018brain,engelbregt2022effects}; meanwhile, its natural sedative effect helps users block out noise, focus their attention, and markedly shorten the time it takes to fall asleep—earning it the nickname “audio sleeping pill” among listeners~\cite{hardian2020improvement,wang2020research}. Furthermore, the practice serves as a non-clinical coping mechanism, providing temporary symptom relief for individuals suffering from chronic pain and depression~\cite{cash2018expectancy,kim2024effects}.

Existing attempts to synthesize whispered or ASMR speech primarily fall into three categories: In-context learning via large-scale prompt-based models \cite{du2024cosyvoice, zhang2025minimax}, Voice Conversion (VC) techniques that transform modal speech into whispers \cite{cotescu2019voice}, and task-specific fine-tuning on limited ASMR datasets \cite{hu2021whispered}. However, these approaches face two critical limitations. The first is poor zero-shot generalization: most systems require an ASMR-style reference from the target speaker. They struggle to generate ASMR speech for a speaker whose only available data is in a normal, voiced speaking style.
The second is unsatisfactory acoustic authenticity: existing models often treat whispering as a simple "style transfer" or "noise addition" process, failing to capture the intricate interplay between breath sounds and unvoiced linguistic content that defines a true ASMR experience.

Therefore, we innovatively introduce DeepASMR, the first framework enabling controllable synthesis of high-quality, personalized ASMR speech from arbitrary text and any speaker's voice.  Our approach leverages a two-stage architecture: a Large Language Model (LLM) based text-to-semantic encoder and a flow-matching-based acoustic decoder. A critical insight of our work is the identification of a latent factorization within the token space, which allows a two-stage model to soft factorize "speaker identity" from "ASMR style" in each stage. This enables the synthesis of ASMR speech for any content and any speaker, even in the absence of their whispered reference samples.

Our main contributions are summarized as follows:

1. \textbf{Zero-Shot Framework}: We present DeepASMR, the first framework specifically formulated to address zero-shot ASMR generation, enabling high-quality synthesis without requiring prior ASMR recordings from the target speaker.

2. \textbf{Token-level Factorization Strategy}: We propose a two-stage model that leverages the inherent soft factorization in the semantic tokens, which primarily encodes ASMR stylistic patterns while retaining residual speaker attributes. This structure facilitates hierarchical control over style and timbre in their respective stages.

3. \textbf{Large-Scale Dataset}: We release DeepASMR-DB, a 670-hour English-Chinese ASMR speech corpus with 35 speakers. To our knowledge, this is currently the largest publicly available ASMR dataset.

4. \textbf{Comprehensive Evaluation}: We introduce a novel evaluation methodology integrating objective metrics, human listening tests, LLM-based scoring and unvoiced speech analysis to overcome the limitations of conventional metrics in assessing ASMR quality.

5. \textbf{State-of-the-Art Performance}: Extensive experiments demonstrate that DeepASMR achieves superior performance in ASMR generation while remaining competitive in conventional speech synthesis tasks. We invite you to listen to our audio samples~\footnote{https://vivian556123.github.io/deepasmr-demo/}.

\section{Related Work}

\subsection{LLM based Text-to-Speech}
The advent of Large Language Models (LLMs) has profoundly reshaped natural language processing, now extending their remarkable capabilities to speech generation. Unlike traditional TTS systems that often rely on complex acoustic models and hand-crafted features~\cite{ren2020fastspeech, wang2017tacotron}, LLM-based approaches represent a paradigm shift by redefine speech synthesis as a token prediction task, where the LLM learns to generate acoustic tokens or semantic codes from text~\cite{borsos2023soundstorm}. 

Diverse approaches have emerged within this paradigm. Some integrate foundational LLMs with existing speech models~\cite{hao2025boosting}, while others directly fine-tune text-based LLMs for speech tasks~\cite{fang2024llama, ye2025llasa}. A prominent trend involves two-stage TTS systems that couple an LLM with an acoustic decoder~\cite{du2024cosyvoice,yang2025emovoice,zhang2024covomix}. The cornerstone of these two-stage architectures is codec, such as X-Codec~\cite{ye2025codec}, S3 Tokenizer~\cite{du2024cosyvoice}, and FACodec~\cite{ju2024naturalspeech}. These tokens are crucial because they can, to some extent, disentangle semantic content from acoustic attributes, offering a flexible representation of speech. This inherent factorization makes the two-stage structure  advantageous for tasks like emotion and style transfer~\cite{yang2025emovoice, zhou2025indextts2}.

\subsection{ASMR Speech Conversion}

Voice conversion (VC) is the task of transforming an individual’s voice into another,
while preserving the linguistic and prosodic content~\cite{liu2025e2e}. Whisper-to-Normal voice conversion, a special case of VC, holds
great promise for assistive communication and healthcare, attracting the attention of many researchers. A common approach is to use self-supervised speech representation learning techniques to extract content information of whisper speech from pre-trained  networks~\cite{rekimoto2023wesper,avdeeva2024improvement}. Other works have designed alternative network structures to reduce training loss and improve inference efficiency, such as cycle-consistent generative adversarial networks~\cite{wagner2023vocoder} and convolutional neural networks~\cite{tan2025distillw2n}.

In contrast, the Normal-to-Whisper voice conversion task lacks sufficient research, with only a few early, naive works, such as digital signal processing based methods~\cite{uchida2021practical} and approaches~\cite{cotescu2019voice} based on GMM and DNN.

From another perspective, recent works have introduced conditional flow matching techniques to decouple style and timbre information, enabling controllable zero-shot voice conversion~\cite{yao2025stablevc,wang2025discl}. These approaches hold potential in addressing the aforementioned challenges.

\subsection{ASMR Speech Synthesis}

Efforts to synthesize ASMR-style speech have traditionally followed two paths. 

The first path to generate whisper-style speech is to fine-tune the model with a small amount of target speaker's whisper-style speech~\cite{hu2021whispered}. While this method effectively generates whisper speech for known speakers, it does not support unseen speakers.

Secondly, zero-shot TTS models using in-context learning techniques have provided new opportunities for ASMR speech synthesis tasks~\cite{du2024cosyvoice,chen2024f5,wang2024maskgct}.
In-context learning is a technique where models adapt to new tasks or generate outputs based on patterns inferred directly from the prompt, without requiring explicit retraining. This allows models to produce high-quality, style-consistent speech when given a relevant prompt, such as generating ASMR-style audio from an ASMR prompt. However, it does not support style conversion when the prompt is from a different speech style.

Furthermore, recent commercial models, such as ElevenLabs~\cite{elevenlabs} and MiniMax~\cite{zhang2025minimax}, are capable of generating speech with astonishing naturalness. However, they only generate speech based on a predefined speaker's voice, and their synthesis is primarily based on vocal fold vibration patterns, which restricts their ability in special style like ASMR.

While zero-shot TTS has matured for voiced speech, the transformation of modal speech into unvoiced, ASMR-style audio remains an unaddressed challenge. DeepASMR, our proposed method, is to the best of our knowledge the first to synthesize ASMR speech from any given input speech.

\section{Methodology}
\subsection{Task Formulation}
\label{sec:task_formulation}
The primary objective of this work is to enable precise control over speaking style (Normal vs. ASMR) while consistently preserving speaker identity, regardless of the input style. Specifically, we define the problem of controllable speech synthesis as a mapping function 
\begin{equation}
\text{DeepASMR}: (T, P_{task}, P_{spk}) \rightarrow \hat{Y}
\end{equation}
where $T$ represents the input text sequence, $P_{task}$ denotes the task prompt,  and $P_{spk}$ serves as the speaker prompt containing identity information. We denote $S_{input}$ and $S_{output}$ as the style (e.g., ASMR or Normal) of the input $P_{spk}$ and the output $\hat{Y}$. The goal is to generate a waveform $\hat{Y}$ that faithfully reflects the linguistic content of $T$ and the prosodic style of $S_{output}$, while maintaining the timbre identity of $P_{spk}$.

We categorize the operational scope of DeepASMR into two primary classes, comprising four distinct sub-tasks: 

\begin{enumerate}
    \item  \textbf{Intra-Style Synthesis ($S_{input} = S_{output}$)}: This class encompasses scenarios where the input and output styles are identical. It includes Normal-to-Normal (N2N) synthesis, serving as a baseline to validate DeepASMR’s general fidelity, and ASMR-to-ASMR (A2A) synthesis, which assesses the framework's capability within the specialized ASMR domain.
    \item \textbf{Cross-Style Synthesis ($S_{input} \neq S_{output}$)}: This class focuses on complex style transfer where input and output styles differ. It features ASMR-to-Normal (A2N) conversion and, most critically, Normal-to-ASMR (N2A) conversion. The latter represents our core contribution: enabling zero-shot ASMR synthesis for a speaker using only their normal, read-style recordings.
\end{enumerate}

\subsection{Framework Overview}

\begin{figure*}[t]
\centering
\includegraphics[width=0.93\textwidth]{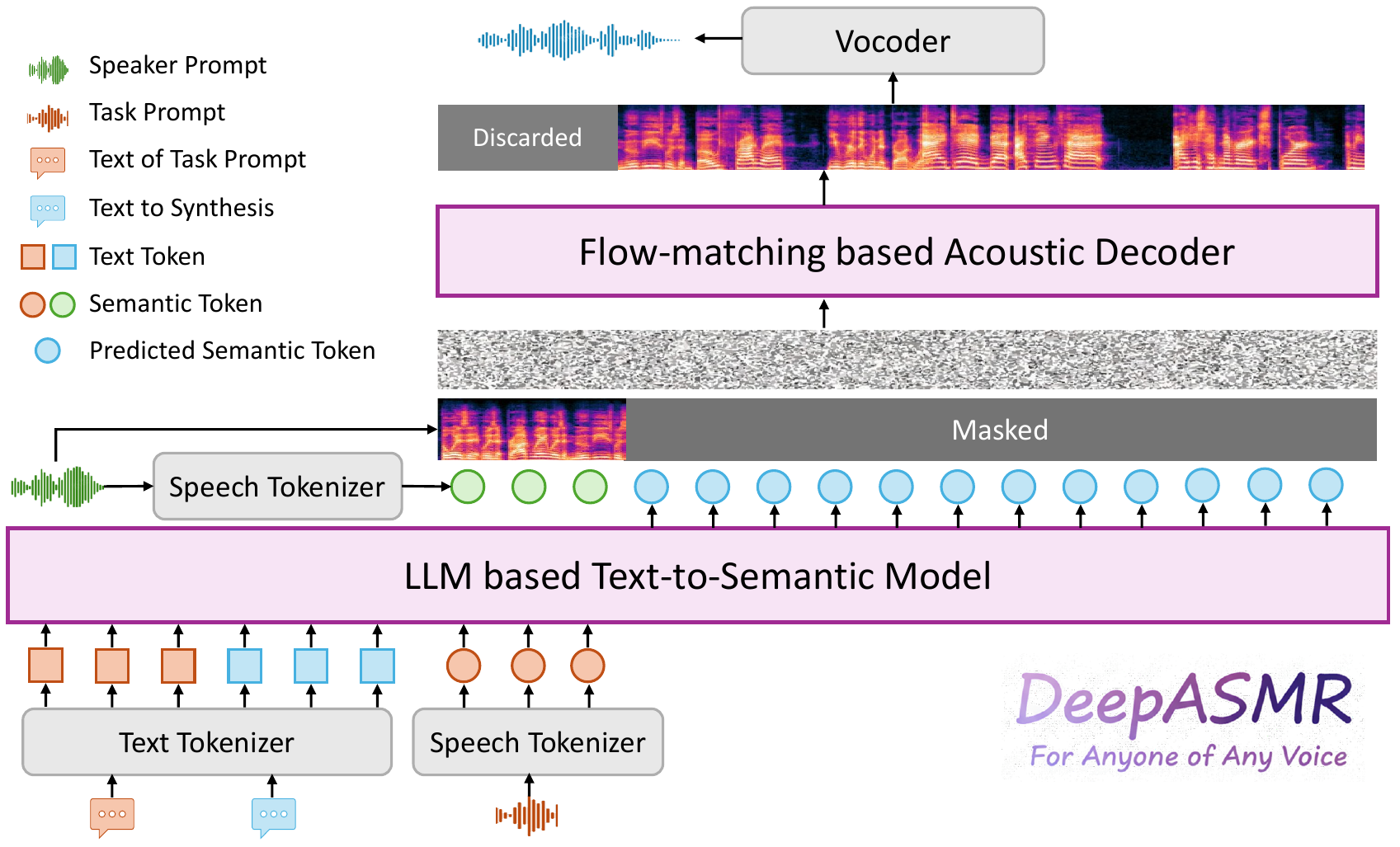}
\caption{DeepASMR Framework Overview}
\label{fig:model}
\end{figure*}

The DeepASMR framework, illustrated in Figure~\ref{fig:model}, is designed for precise speech style control and speaker identity preservation, without requiring a style-matched sample from the target speaker.  DeepASMR adopts a two-stage generative pipeline, decoupling semantic modeling from acoustic reconstruction.

The first stage is a LLM-based Text-to-Semantic Model. We employ a decoder-only Large Language Model (LLM) initialized with Qwen2.5-0.5B~\cite{qwen2025qwen25technicalreport} pre-trained weights. This module functions as a content-style encoder. The input sequence is constructed by concatenating the Task Prompt $P_{task}$ with the target style $S_{output}$ and the Target Text $T$. The LLM autoregressively predicts a sequence of discrete speech tokens $\mathbf{z} = \{z_1, z_2, ..., z_T\}$ from a specific codebook. We utilize S3 tokens~\cite{du2024cosyvoice}, derived from a tokenizer trained with an Automatic Speech Recognition (ASR) objective. The model is optimized using the standard Cross-Entropy loss in Equation~\ref{eq:cross_entropy_loss} over the predicted token sequence: 
\begin{equation}
\label{eq:cross_entropy_loss} 
\mathcal{L}_{CE} = - \sum \log P(z_t | z_{<t}, T, P_{task})
\end{equation}

The second stage is the flow-matching acoustic decoder, which reconstructs the acoustic details (mel-spectrogram) from the predicted semantic tokens. We implement a Conditional Flow Matching network following Voicebox~\cite{le2023voicebox} based on a transformer encoder architecture~\cite{le2023voicebox, eskimez2024e2} with loss in Equation \ref{eq:cfm}, where $x_1$ is the ground truth Mel-spectrogram, $x_0$ is a sample of Gaussian noise, and $\sigma_{min}$ is the minimum noise level. The decoder is conditioned on two inputs: the semantic token sequence $\mathbf{z}$ (providing content and gross prosody) and the mel-spectrogram of the Speaker Prompt $P_{spk}$ (providing fine-grained timbre). The generated mel-spectrogram is finally converted into a time-domain waveform using a pre-trained HiFi-GAN vocoder~\cite{kong2020hifi}.
\begin{equation}
\label{eq:cfm}
    \mathcal{L}_{CFM} = \mathbb{E}_{t, x_1, x_0} [ || v_t(x_t, \mathbf{z}) - (x_1 - (1-\sigma_{min})x_0) ||^2 ]
\end{equation}

\subsection{Analysis of Token-level Soft Factorization}
\label{sec:tokenizer}

A core premise of our approach is that discrete speech tokens enable a soft factorization of style from timbre and are predominantly shaped by style rather than timbre. We validate this hypothesis through rigorous analysis of the S3 tokenizer~\cite{du2024cosyvoice}.

The S3 tokenizer utilizes Finite Scalar Quantization (FSQ)~\cite{mentzer2023finite} trained on an ASR task. This quantization acts as an information bottleneck: it retains linguistic content and macro-prosodic features (such as the slow speaking rate and distinct pauses typical of ASMR) while discarding most of the high-frequency speaker-specific details (such as exact harmonic structures). 

\begin{figure}[t]
\centering
\includegraphics[width=0.99\columnwidth]{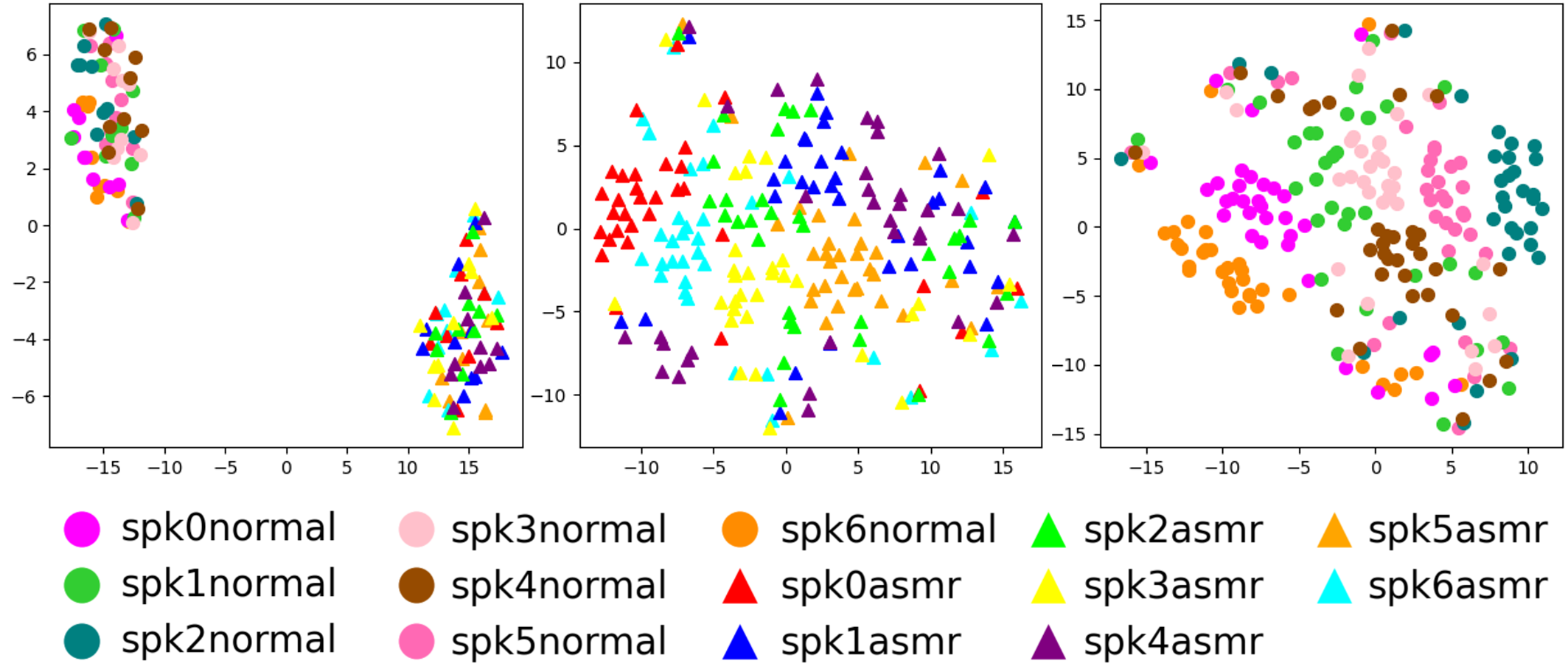}
\caption{tSNE visualization of speaker embedding extracted from S3 tokenizer hidden features}
\label{fig:disentangle}
\end{figure}

To empirically validate this factorization property, we extracted pre-quantized hidden states from the tokenizer for a controlled set of 7 speakers (paired Normal/ASMR data). We computed frame-level embeddings via mean-pooling and performed t-SNE dimensionality reduction in Figure \ref{fig:disentangle}. The visualization reveals two critical insights: 

\begin{itemize} 
\item \textbf{Style-Dominant Distribution:} The embedding space shows a distinct hyperplane separation between Normal and ASMR clusters. The variance attributed to style significantly outweighs that of speaker identity, indicating that the token space is predominantly shaped by style. 
\item \textbf{Residual Timbre Information:} Within each style cluster, samples from different speakers exhibit weak but observable grouping tendencies. This confirms that while the tokenizer filters significant acoustic information, it still carries residual timbre information, supporting the notion of soft factorization rather than complete isolation, which necessitates our specific design to prevent identity leakage. 
\end{itemize}

To quantify the extent of speaker information retention, we trained a convolutional neural-network based speaker classifier on the S3 tokenizer's hidden states using the 36 speakers from CHAINs~\cite{cummins2006chains} for 30 epochs. While a baseline classifier trained on Mel-spectrograms achieved 90\% accuracy, the S3-hidden-based classifier dropped to 86.4\%, significantly surpassing the random chance baseline of 2.8\%. This high retention rate quantitatively confirms our soft factorization hypothesis: although the tokenizer compresses the audio, substantial speaker identity persists within the tokens.

These observations justify our dual-modeling strategy: since ASMR involves macro-prosodic features—such as reduced speech rate and elongated vowel durations~\cite{houle2020acoustic} that are inherently embedded in the token sequences—we employ an LLM to model these dominant stylistic variations. Concurrently, a Flow-based decoder is utilized to reconstruct the residual acoustic details and the subtle speaker-specific information, effectively recovering the complete timbre.

\subsection{Task Prompt Selection via Virtual Speaker Pool}

\label{sec:task_selector}
As shown in Section \ref{sec:task_formulation}, we define two synthesis tasks: intra-style synthesis, which maintains the speaker's original style, and cross-style synthesis, which converts speech to a different style. 

The primary challenge in cross-style synthesis stems from the residual speaker information embedded within speech tokens (as analyzed in Section~\ref{sec:tokenizer}). In cross-style tasks, using an arbitrary ASMR recording as a style prompt often leads to "timbre leakage," where the output voice drifts towards the reference speaker. To resolve this, we propose an automated Task Prompt Selector based on a Virtual Speaker Pool.

\begin{figure}[t]
\centering
\includegraphics[width=0.9\columnwidth]{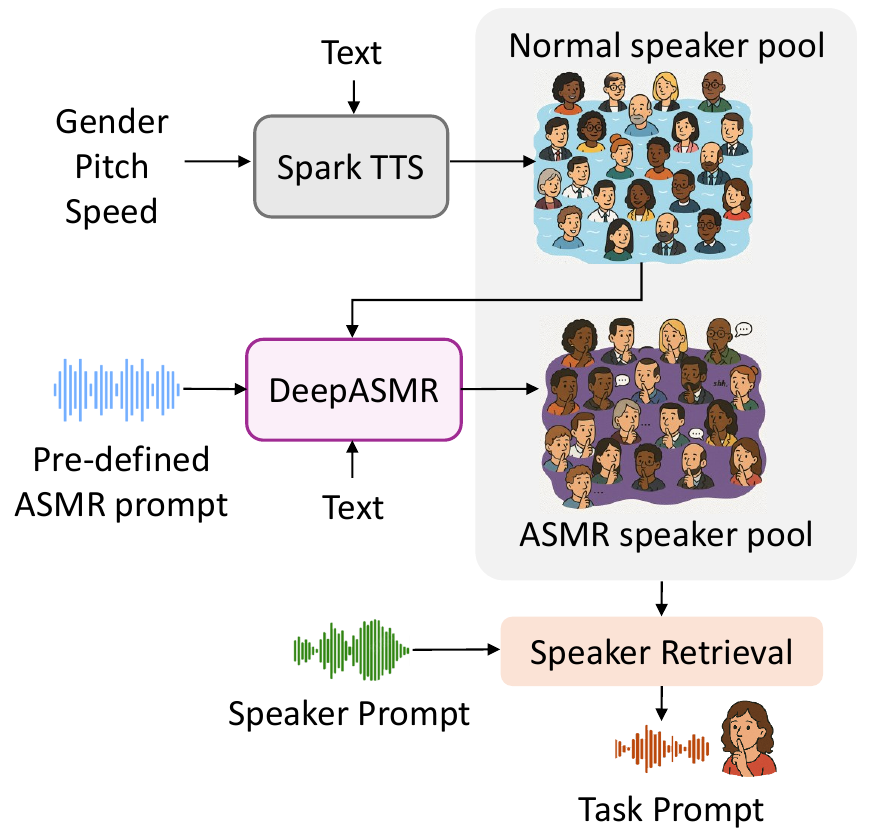}
\caption{Illustration of speaker pool construction and task prompt selection}
\label{fig:speaker_pool}
\end{figure}

We construct a synthetic, controllable speaker space to serve as a robust reference bank and then extract the best task prompt through similarity-based retrieval, as illustrated in Figure \ref{fig:speaker_pool}.

\subsubsection{Pool Construction}

Technically, three strategies exist for prompt selection: (1) Static Prompting, using a single universal reference for all utterances, which risks significant speaker information drift, especially when gender or vocal profiles mismatch; (2) Manual Selection, which involves hand-picking a compatible reference for each target speaker—a process that is precise but lacks scalability for large-scale applications; and (3) Virtual Pool Retrieval, our proposed automated approach. 

To balance synthesis quality with operational efficiency and speaker privacy, we construct two distinct virtual speaker pools—one for normal speech and one for ASMR—each containing a diverse range of synthetic vocal profiles. 

\begin{enumerate}
\item \textbf{Normal Pool ($V_{norm}$):} Consists of 50 utterances $\{u_i^{\text{norm}}\}$  synthesized via SparkTTS~\cite{wang2025spark}. To ensure broad coverage of the acoustic manifold, we systematically vary three dimensions: gender (male and female), pitch ("very\_low", "low", "moderate", "high", "very\_high"), and speed ("very\_low", "low", "moderate", "high", "very\_high"), resulting in a diverse set of vocal profiles. 

\begin{equation}
u_i^{\text{norm}} \sim \text{SparkTTS}(\text{Gender}, \text{Pitch}, \text{Speed})
\end{equation}

\item \textbf{ASMR Pool ($V_{asmr}$):} Contains 50 utterances generated by our DeepASMR model to serve as stylistic counterparts to the Normal Pool. Specifically, each entry is synthesized by using a pre-defined high-quality ASMR sample $u^{\text{asmr}}$ from the training set as the task prompt, while the 50 diverse utterances from the Normal Pool $V_{norm}$ serve as the speaker prompts. This one-to-one mapping ensures that the ASMR pool covers the same breadth of vocal identities as the Normal Pool, but within the target ASMR stylistic domain. 
\end{enumerate}
\begin{equation}
u_i^{\text{asmr}} \sim \text{DeepASMR}(T, u^{\text{asmr}}, u_i^{\text{norm}})
\end{equation}
\subsubsection{Similarity-based Retrieval}
For a given target speaker, we employ WeSpeaker as a pre-trained speaker verification system~\cite{wang2023wespeaker} to extract the speaker embedding $e_{tgt}$ and the speaker embedding for all candidates $\{e_i\}_{i=1}^{50}$ in the Virtual Pool. We compute the cosine similarity between the $e_{tgt}$ and  $\{e_i\}_{i=1}^{50}$  of all candidates. For example, the candidate with the maximum similarity is selected as the optimal task prompt for the ASMR generation following Equation \ref{eq:retrieval}.

\begin{equation}
\label{eq:retrieval}
P_{task} = u^{asmr}_{k} \quad \text{where} \quad k = \arg \max_{i} \text{CosSim}(e_{tgt}, e_i)
\end{equation}

By selecting a vocal  neighbor as the task prompt, we ensure that the LLM focuses on modeling the target style variations while the residual speaker information remains consistent with the target identity, leading to a more coherent and high-fidelity synthesis.

\begin{figure*}[t]
\centering
\includegraphics[width=0.95\textwidth]{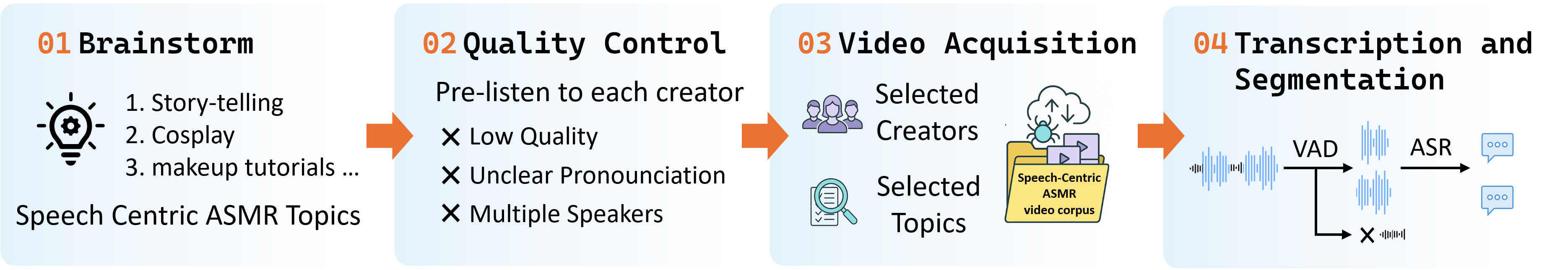}
\caption{A comprehensive pipeline for constructing the DeepASMR-DB dataset}
\label{fig:data}
\end{figure*}

\subsection{Training and Inference Pipeline}
\label{sec:training-and-inference}
\subsubsection{Two-Stage Training} DeepASMR's training is a two-stage sequential process, building robust Text-to-Speech (TTS) capabilities before specializing in ASMR generation. 

We first train the LLM and Flow Decoder on 200k hours of internal TTS data (80k Chinese data and 120k English data) for 250k steps. 
To specialize in ASMR while retaining stability, we fine-tune on a mixture of DeepASMR-DB (our ASMR speech dataset) and the Emilia dataset (normal speech dataset)~\cite{he2024emilia} for 40 epochs. The mixing strategy is crucial to prevent the model from overfitting to whispered speech and losing its ability to generate voiced phonemes. 

\subsubsection{Inference Process} Let $T$ be the input text and $P_{spk}$ be the target speaker's prompt recording. The task prompt representing the target style is determined as follows, where $\mathcal{F}_{\text{retrieval}}$ denotes the similarity-based retrieval from the Virtual Pool $\mathcal{V}$ in Section \ref{sec:task_selector}.
\begin{equation}
\label{eq:task_selector}
P_{task} =
  \begin{cases}
  P_{spk} & \text{if Intra-Style } \\
  \mathcal{F}_{\text{retrieval}}(P_{spk}, \mathcal{V}) & \text{if Cross-Style }
  \end{cases}
  \end{equation}
The LLM operates as a conditional probability distribution, predicting the target semantic tokens $\mathbf{z}_{pred}$ autoregressively, and the acoustic decoder generates the mel-spectrogram $M_{out}$ by modeling the flow field from a Gaussian prior. Crucially, it is conditioned on the concatenated token sequence $[\mathbf{z}_{spk}, \mathbf{z}_{pred}]$ and the fine-grained timbre features of the original speaker prompt. Finally, the waveform is synthesized via a HiFi-GAN vocoder~\cite{kong2020hifi}.

\subsubsection{Iterative Inference Refinement} 
\label{sec:repeating}
For challenging cross-style cases, we employ an iterative refinement strategy. The output of the first pass (N2A generation) can be fed back into the system as a new, synthetic speaker prompt $P_{spk}$. As shown in our ablation study (Table \ref{tab:ablation_study_repeat}), repeating this inference process 2-3 times can further enhance the unvoiced whispering speech quality by reinforcing the style features, although a trade-off exists with speaker similarity.

\section{DeepASMR-DB}

\subsection{Overview of DeepASMR dataset}
We introduce DeepASMR-DB, a large-scale, high-fidelity bilingual dataset specifically curated to advance research in Autonomous Sensory Meridian Response (ASMR) speech synthesis and analysis. Unlike standard speech corpora that focus on neutral or emotive prosody, ASMR necessitates a unique acoustic profile characterized by low-velocity vocalizations, intricate breath patterns, and intimate microphone proximity

DeepASMR-DB addresses the current scarcity of high-quality ASMR data by providing over 40,000 meticulously transcribed samples, totaling 674.5 hours of audio. To our knowledge, this represents the largest bilingual, multi-speaker dataset of its kind. The dataset is balanced across two major languages, comprising 141.7 hours of Mandarin Chinese (22 native speakers) and 532.8 hours of English (13 native speakers). It features 35 distinct speakers (28 female, 7 male), ensuring a rich diversity of vocal textures and reflects the female-dominated demographics of the ASMR creator community.

To guarantee robustness for real-world applications, we curated content across nine distinct categories, ranging from structured poem readings and scientific expositions to spontaneous formats such as game commentaries and role-play scenarios. This topical breadth allows the model to learn not only the whisper-style phonation characteristic but also the linguistic nuances required for varied conversational contexts. 

The dataset is under a CC BY-NC 4.0 license. The copyright of the audio content remains with the original creators. Audio samples are provided for review, and the complete dataset will be made publicly available upon the paper's acceptance~\footnote{https://github.com/vivian556123/DeepASMR-DB-samples}.

\subsection{Data selection and preparation pipeline}

The construction of DeepASMR-DB followed a rigorous four-stage pipeline designed to balance acoustic purity with the unique stylistic requirements of ASMR content, as illustrated in Figure \ref{fig:data}. 

\subsubsection{Brainstorm and Topic-Driven Filtering} The process began with a strategic topic-driven filtering phase, where we identified and prioritized video categories that are inherently speech-dependent. The primary motivation for this step is to decouple meaningful linguistic content from non-verbal auditory triggers—such as tapping, scratching, or crinkling—which are prevalent in ASMR recordings. While these triggers are essential to the ASMR experience, they lack phonetic information and fall outside the scope of conventional ASR systems; their presence often introduces significant noise that can degrade recognition accuracy or mislead the model. 

Therefore, to improve data source selection, we employed more fine-grained keyword indexing. We selected nine topics, such as storytelling, question-and-answer sessions, cosplay narratives, cosmic makeup tutorials, poem reading, scientific and historical introductions, picture description, game commentaries, and personal perspectives, while excluding keywords like eating and ear-spa to avoid the inclusion of excessive non-speech information.

\subsubsection{Quality Control and Creator Selection} Following the initial filtering, we conducted a manual vetting of potential creators across major platforms such as YouTube and Bilibili. Since a creator's articulation, recording equipment, and acoustic environment remain largely consistent across different videos, our first step is to assess whether the creator's speech quality meets our requirements. At this stage, we exclusively retain single-speaker content and specifically select creators who exhibit clear articulation and maintain consistent low-vibration or non-vibrato vocalizations—the defining hallmarks of the ASMR experience. This process involves manual annotation and verification. A creator is only selected if at least 80\% of their randomly sampled videos (10 segments per creator) meet the predefined quality requirements. 

During this phase, we prioritized acoustic fidelity and professional articulation over strict demographic balancing. Consequently, the final selection exhibits a gender imbalance (28 female vs. 7 male) that mirrors the female-dominated demographics of the high-quality ASMR creator community.

This high-standard vetting process was essential to ensure that the dataset remains professionally high-fidelity and provides a stable baseline of ASMR-specific prosody, preventing the inclusion of erratic volume spikes or poor-quality recordings that could degrade model performance.

\subsubsection{Video Acquisition}Once the creators and topics were finalized, we proceeded with the systematic acquisition of all relevant audio files. Specifically, we downloaded all videos under the target topics for each selected creator to build a statistically significant corpus.

\subsubsection{Transcription and Segmentation} The final stage involved an automated transcription and segmentation workflow. Given the excessive length of the original videos, we first segmented them into 10-minute intervals and subsequently utilized the Microsoft Fast Transcription API~\footnote{https://learn.microsoft.com/en-us/azure/ai-services/speech-service/fast-transcription-create} for annotation. 
 A critical advantage of this annotation phase was the selective preservation of non-verbal acoustic cues. While it discarded pure long silence, it deliberately retained breath sounds and subtle, speech-accompanied elements inherent to the ASMR genre. This approach ensures the dataset captures the intimate, high-frequency textures necessary for authentic ASMR synthesis—details that standard speech corpora typically filter out.

\section{Experimental Setups}
\subsection{Training Configuration}
During the training process, similar to CosyVoice2~\cite{du2024cosyvoice}, the DeepASMR LLM is initialized with the pre-trained Qwen2.5-0.5B model~\cite{qwen2025qwen25technicalreport}. The acoustic decoder is implemented as a Transformer encoder, with 24 layers, 16 attention heads, and an embedding dimension of 1024 with U-Net~\cite{unet} style skip connections. 

Training proceeds in two stages. In the first training stage, the model is pre-trained on 200,000 hours of internal data for 250k steps. The Adam optimizer is employed for both LLM and acoustic model, with 10,000 warm-up steps and a Noam learning rate scheduler~\cite{vaswani2017attention}, where the peak learning rate is set to $1e-4$ for LLM and 1e-5 for acoustic decoder. In the second training stage, both LLM and acoustic decoder are fine-tuned using 1,000 hours of data extracted from the Emilia~\cite{he2024emilia} and our DeepASMR-DB for an additional 10 and 40 epochs for the LLM and acoustic model respectively. For this stage, a constant learning rate of $1e-5$ is used. All experiments are conducted on eight NVIDIA A100 GPUs. We set gradient accumulation to 2 and use a dynamic batcher with 23000 frames per batch. 

\subsection{Evaluation Metrics}
We adopt a multi-dimensional evaluation protocol comprising objective, subjective, and LLM-based metrics to comprehensively assess performance.

\subsubsection{Objective Metrics} Following SeedTTS, we utilize Whisper~\cite{radford2023robust} to measure the Word-Error-Rate (WER), and Paraformer~\cite{gao2022paraformer} to measure Character-Error-Rate (CER) for the Chinese test set of the generated speech to assess intelligibility. Speaker similarity (SIM) is evaluated using WavLM-Large~\cite{chen2022wavlm}, measuring the cosine similarity between the embeddings of the generated speech and the prompt speech.  
 
  \subsubsection{Subjective Metrics} We conducted a Mean Opinion Score (MOS) test with 12 human listeners. Participants evaluated samples based on two criteria: Overall Impression (OI-MOS), assessing general audio quality and naturalness, and ASMR-Specific Comfort (ASMR-MOS), quantifying the relaxation and tingling sensation characteristic of the genre.
 
  \subsubsection{LLM-based Metrics} Given the limitations of standard metrics in capturing stylistic nuance, we utilized the Gemini 2.5 Pro model~\cite{comanici2025gemini} for automated style detection. Through prompt engineering, the model scores speech style on a continuous scale from $[-1, +1]$, where $-1$ indicates a normal style, $+1$ indicates an ASMR style, and $0$ denotes an ambiguous or unknown style. We release the detailed instruction of our LLM-based evaluation\footnote{https://github.com/ztxiao1211/DeepASMR-testset}.
 
 \subsubsection{Unvoiced Speech Metrics} 
To rigorously evaluate the model's ability to generate unvoiced speech, we employed a frame-level acoustic analysis to distinguish between silence, voiced speech, and unvoiced speech in the challenging Normal-to-ASMR task. We calculate the Global Unvoiced Ratio ($R_{UV}$) using a two-stage classification process based on energy and periodicity.  We analyze the generated waveforms using a frame-by-frame approach to extract RMS energy ($E$) and fundamental frequency ($f_0$) using the probabilistic YIN (PYIN) algorithm~\cite{mauch2014pyin}. Following ~\cite{guidi2017features}, frames are classified into three categories:
\begin{itemize}
    \item Silence: Frames where energy falls below an adaptive threshold ($E_i < 0.02 \cdot \max(E)$).
    \item Voiced Speech: Active frames where a valid $f_0$ is detected.
    \item Unvoiced Speech ($U_i$): Active frames where no periodic component is detected ($f_0$ is undefined/NaN).
\end{itemize}
The Global Unvoiced Ratio is defined as the percentage of active speech frames that are unvoiced. 
\begin{equation}
R_{UV} = \frac{\sum U_i}{\sum A_i} \times 100\%
\end{equation}
where $U_i$ is the binary indicator for unvoiced frames and $A_i$ is the binary indicator for all active speech frames (voiced + unvoiced). A score approaching $100\%$ indicates the successful generation of unvoiced speech without periodic leakage. 

\subsection{Baselines}
We benchmark DeepASMR against leading zero-shot synthesis models across two synthesis categories:

For Intra-Style Synthesis ($S_{input} = S_{output}$), we selected two main-stream zero-shot TTS models—CosyVoice2~\cite{du2024cosyvoice} and F5TTS~\cite{chen2024f5}—as baseline models. These models utilize in-context learning techniques, which demonstrate strong generalization across various speech styles and show the ability to generate high-quality ASMR-style speech. We also fine-tune these models with our proposed DeepASMR-DB to have a fair comparison on the ASMR generation scenario. The generated speech can be formulated as a mapping as follows:
\begin{equation}
\hat{y} = \mathcal{F}_{TTS}(T, P_{spk})
\end{equation}

However, for the Cross-Style Synthesis ($S_{input} \neq S_{output}$) tasks, to the best of our knowledge, there is no directly prior work. Standard zero-shot models lack the disentanglement capability to map a prompt $P_{spk}$ with source style $S_{input}$ (e.g., Normal) to a target style $S_{output}$ (e.g., ASMR). To address this, we formulate a cascade baseline combining Text-to-Speech (TTS) and Voice Conversion (VC) as follows:

\begin{enumerate}
\item Auxiliary Prompt Selection: We utilize a fixed auxiliary speaker $P_{aux}$ which possesses the target style $S_{output}$ but comes from a different speaker. 
    \item Style Injection: We generate an intermediate waveform $y_{inter}$ using a TTS model conditioned on the auxiliary prompt. This yields speech with the correct content $T$ and target style $S_{output}$, but the auxiliary speaker's timbre.
    \begin{equation}
    y_{inter} = \mathcal{F}_{TTS}(T, P_{task})
    \end{equation}
    \item Timbre Transfer: We apply a Voice Conversion model $\mathcal{F}_{VC}$ to shift the timbre from $P_{aux}$ to the target speaker $P_{spk}$, while attempting to preserve the style in $y_{inter}$.
    \begin{equation}\hat{y} = \mathcal{F}_{VC}(y_{inter}, P_{spk})
    \end{equation}
\end{enumerate}

For the VC component $\mathcal{F}_{VC}$, we selected two main-stream VC models, CosyVoiceVC~\cite{du2024cosyvoice} and SeedVC~\cite{plachtaa}, creating a two-stage pipeline to approximate the DeepASMR objective.


\subsection{Testing Dataset}
To rigorously evaluate DeepASMR across linguistic and stylistic dimensions in unseen speakers, we constructed a comprehensive evaluation benchmark comprising eight distinct test sets (2 languages $\times$ 4 sub-tasks).

To ensure that performance differences are attributable to style modeling rather than speaker identity or linguistic variation, we selected datasets that provide paired Normal and ASMR recordings for the same speakers, allowing for a direct, controlled comparison between Intra-Style and Cross-Style synthesis tasks. 
We utilize CHAINS corpus~\cite{cummins2006chains} for English testing set and Whisper40 dataset~\cite{yang2024whisper40} for chinese testing set. By leveraging these paired datasets, we establish a ground truth for both timbre and style in all scenarios, enabling the precise objective measurement of style transfer fidelity. The detailed test set configuration is available online\footnote{https://github.com/ztxiao1211/DeepASMR-testset}. 

\section{Results and Analysis}

\subsection{Objective Evaluation}

\begin{table*}[t]
    \centering
      \setlength{\tabcolsep}{5pt}
        \begin{tabular}{c|c|cc|c|cc|c|cc|c|cc|c}
        \toprule
        \multicolumn{14}{c}{\textit{Intra-style Synthesis}} \\ \midrule
       \multicolumn{2}{c|}{\multirow{3}{*}{TTS Model}}  & \multicolumn{6}{c|}{Normal $\rightarrow$ Normal} & \multicolumn{6}{c}{ASMR $\rightarrow$ ASMR} \\
       \multicolumn{2}{c|}{} & \multicolumn{3}{c|}{ZH} & \multicolumn{3}{c|}{EN} & \multicolumn{3}{c|}{ZH} & \multicolumn{3}{c}{EN} \\
        \multicolumn{2}{c|}{} & CER $\downarrow$ & SIM $\uparrow$ & Style $\downarrow$ & WER $\downarrow$ & SIM  $\uparrow$& Style $\downarrow$ & CER $\downarrow$ & SIM $\uparrow$ & Style  $\uparrow$ & WER $\downarrow$& SIM $\uparrow$ & Style  $\uparrow$ \\
        \midrule
        \multicolumn{2}{c|}{Ground Truth} & 4.38 & 0.82 & \textbf{-1.00} & 3.00 & 0.70 & -0.96 & 43.22 & 0.75 & +0.65 & 5.11 & 0.58 & +0.42 \\ \midrule
\multicolumn{2}{c|}{CosyVoice2} & 4.62 & \textbf{0.80} & \textbf{-1.00} & 2.96 & 0.56 & \textbf{-0.98} & 26.53 & 0.72 &  +0.55 & 5.30 & \textbf{0.47} & +0.58 \\ 
\multicolumn{2}{c|}{CosyVoice2$_{FT}$} & \textbf{3.54} & 0.78 & \textbf{-1.00} & \textbf{1.80} & 0.53 & -0.95 & \textbf{21.55} & 0.70 & +0.79 & \textbf{5.05} & \textbf{0.47} & +0.44 \\ 
\multicolumn{2}{c|}{F5TTS} & 12.52 & \textbf{0.80} & -1.00 & 5.49 & 0.50 & -0.96 & 31.03 & 0.62 & -0.04 & 7.44 & 0.23 & -0.63 \\  
\multicolumn{2}{c|}{F5TTS$_{FT}$} & 13.58 & 0.77 & -1.00 & 5.80 & 0.44& -0.96& 26.76& \textbf{0.73}& +0.68 & 6.67& 0.41& +0.38 \\

\midrule
\multicolumn{2}{c|}{DeepASMR} & 5.58 & 0.79 & -0.97 & 1.91 & \textbf{0.59} & -0.88 & 28.95 & 0.72& \textbf{+0.84} & 6.72 & \textbf{0.47} & \textbf{+0.77} \\ 
        \midrule \midrule
        \multicolumn{14}{c}{\textit{Cross-style Synthesis}} \\  \midrule
        & & \multicolumn{6}{c|}{ASMR $\rightarrow$ Normal} & \multicolumn{6}{c}{Normal $\rightarrow$ ASMR} \\
        TTS Model & VC Model & \multicolumn{3}{c|}{ZH} & \multicolumn{3}{c|}{EN} & \multicolumn{3}{c|}{ZH} & \multicolumn{3}{c}{EN} \\
         & & CER $\downarrow$ & SIM $\uparrow$ & Style $\downarrow$ & WER $\downarrow$ & SIM  $\uparrow$& Style $\downarrow$ & CER $\downarrow$ & SIM $\uparrow$ & Style  $\uparrow$ & WER $\downarrow$& SIM $\uparrow$ & Style  $\uparrow$ \\
        \midrule
\multicolumn{2}{c|}{Ground Truth} & 4.38 & 0.32 & -1.00 & 3.06 & 0.37 & -0.96 & 43.55 & 0.47 & +0.64 & 5.11 & 0.37 & +0.39 \\ \midrule
CosyVoice2 &  \multirow{2}{*}{CosyVoiceVC} & 7.49 & 0.46 & \textbf{-1.00} & 3.37 & 0.33 & \textbf{-0.84} & 36.29 & 0.61 & -0.53 & 11.48 & 0.44 & -0.01 \\ 
F5TTS  & & 14.63 & 0.47 & -0.96 & 9.15 & 0.32 & -0.68 & 23.01 & 0.66 & -0.81 & 10.51 & 0.44 & -0.77 \\ 
\midrule
CosyVoice2 & \multirow{2}{*}{SeedVC}  & 11.66 & 0.35 & -0.41 & 6.68 & \textbf{0.34} & -0.39 & 41.13 & 0.62 & -0.91 & 13.47 & \textbf{0.45} & -0.71 \\ 
F5TTS & & 20.86 & \textbf{0.49} & -0.05 & 12.34 & 0.32 & -0.48 & 30.97 & \textbf{0.63} & -0.93 & 15.28 & 0.41 & -0.89 \\  
\midrule
\multicolumn{2}{c|}{DeepASMR} & \textbf{3.34} & 0.46 & -0.50 & \textbf{1.90} & 0.31 & -0.82 & \textbf{19.29} & 0.58 & \textbf{+0.65} & \textbf{6.53} & 0.41 & \textbf{+0.58} \\ 
    \bottomrule
    \end{tabular}%
    \caption{Objective and LLM-based Evaluation Results for Different Models under Intra and Cross-Style Synthesis Scenarios}
    \label{tab:objective}
\end{table*}

We initiated our analysis by evaluating the four defined sub-tasks using conventional TTS objective metrics: Word Error Rate (WER) or Character Error Rate (CER) for intelligibility, and Speaker Similarity (SIM) for timbre preservation. Additionally, given the limitations of standard metrics in capturing prosody, we incorporated an LLM-based Style score (ranging from -1 for Normal to +1 for ASMR) to quantify stylistic fidelity.

In terms of intelligibility, as presented in Table \ref{tab:objective}, DeepASMR demonstrates exceptional robustness in cross-style synthesis scenarios. In the challenging Normal-to-ASMR task, our model achieved the lowest error rates across both languages (e.g., 6.53\% WER in English), significantly outperforming the cascade Voice Conversion baselines, which suffered from high degradation. In intra-style synthesis, DeepASMR remained highly competitive. While the fine-tuned CosyVoice2 model achieved the marginally lowest error rates in specific cases, our DeepASMR model consistently outperformed the zero-shot F5TTS baseline and maintained stability comparable to the ground truth. 

In terms of style conversion and timbre preservation, a critical observation from Table \ref{tab:objective} is the decoupling of style and timbre in cross-style tasks. In the Normal-to-ASMR task, all models exhibited lower SIM scores compared to intra-style tasks. However, comparing these against the ground truth reveals that this drop is more like an intrinsic property of the metric than a model failure. The ground truth itself shows a SIM score of only 0.47 (ZH) and 0.37 (EN) when comparing a speaker's ASMR recordings to their own normal speech. DeepASMR matches this theoretical upper bound almost perfectly, indicating it preserves the maximum amount of speaker identity possible after a radical style shift. Conversely, the VC baselines failed to achieve effective style conversion. Their LLM-based Style scores remained negative, indicating the output remained closer to a normal style rather than the target ASMR. DeepASMR was the only model to achieve high positive Style scores in cross-style synthesis, closely aligning with the stylistic intensity of the ground truth.

Moreover, to validate the quality of our proposed DeepASMR-DB and isolate the impact of our architectural design, we benchmarked DeepASMR against baseline models (CosyVoice2 and F5TTS) explicitly fine-tuned on this dataset. The significant performance gains observed in the fine-tuned variants confirm the high fidelity of DeepASMR-DB: for instance, fine-tuning F5TTS on our corpus drastically improved its English ASMR style score from a normal $-0.63$ to a ASMR $+0.38$. However, despite training on this data, DeepASMR consistently surpassed these fine-tuned specialists in stylistic fidelity. This indicates that DeepASMR-DB provides the necessary acoustic features for ASMR synthesis, and our framework's specialized factorization strategy maximizes stylistic authenticity—performance compared with general models.

\subsection{Subjective Evaluation}
Table \ref{tab:human} summarizes the subjective evaluation results derived from a listening test with 12 human participants. To provide a granular assessment of performance, we employed two distinct metrics: Overall Impression MOS (OI-MOS), which evaluates general acoustic quality, intelligibility, and naturalness; and ASMR-Specific Comfort MOS (ASMR-MOS), which specifically quantifies the relaxation response and "tingling" sensation (autonomous sensory meridian response) characteristic of the genre. 

In the intra-style synthesis scenario, DeepASMR demonstrated superior performance in generating authentic ASMR content. It consistently achieved the highest ASMR-MOS scores, validating the effectiveness of our specialized method in capturing the subtle, breathy and unvoiced acoustics required for relaxation. Among the baselines, CosyVoice2 proved to be the strongest competitor, significantly outperforming other systems like F5TTS in preserving ASMR textures. Importantly, DeepASMR's performance on normal speech (OI-MOS) remained comparable to leading general-purpose TTS models. This confirms that our model's specialization in the ASMR domain does not compromise its ability to synthesize high-fidelity modal speech, ensuring robust performance across diverse stylistic requirements.

The advantages of DeepASMR became most pronounced in cross-style tasks, especially in the challenging Normal-to-ASMR task. DeepASMR achieved a dominant ASMR-MOS of 3.99 in Chinese and 3.91 in English, drastically outperforming the cascade baselines. This validates that only DeepASMR successfully induces the requisite relaxation response when converting from a normal voice. An analysis of the SeedVC baseline reveals a distinct trade-off. While it achieved a moderate OI-MOS (4.08) compared to its very low ASMR-MOS (2.12) in English tasks, this is because it produces clear "whisper-like" normal speech rather than authentic ASMR. Listeners rated it higher on general impression due to audio clarity but penalized it on ASMR-MOS for lacking the true "tingling" texture and unvoiced comfort. DeepASMR, in contrast, balanced both, achieving the highest OI-MOS and ASMR-MOS among all synthesized systems, closely approaching the ground truth.

\begin{table*}[t]
 \setlength{\tabcolsep}{7pt}
    \centering
        \begin{tabular}{c|c|c|c|cc|cc}
        \toprule
        \multicolumn{8}{c}{\textit{Intra-style Synthesis}} \\ \midrule
       \multicolumn{2}{c|}{\multirow{3}{*}{TTS Model}}  & \multicolumn{2}{c|}{Normal $\rightarrow$ Normal} & \multicolumn{4}{c}{ASMR $\rightarrow$ ASMR} \\
       \multicolumn{2}{c|}{} & \multicolumn{1}{c|}{ZH} & \multicolumn{1}{c|}{EN} & \multicolumn{2}{c|}{ZH} & \multicolumn{2}{c}{EN} \\
        \multicolumn{2}{c|}{} & OI-MOS & OI-MOS & OI-MOS & ASMR-MOS & OI-MOS & ASMR-MOS  \\
        \midrule
        \multicolumn{2}{c|}{Ground Truth} & 3.84 $\pm$ \footnotesize{1.16} & 4.05 $\pm$ \footnotesize{1.22} & 4.29 $\pm$ \footnotesize{0.66} & 4.26 $\pm$ \footnotesize{0.67} & 4.38 $\pm$ \footnotesize{0.65} & 4.07 $\pm$ \footnotesize{0.84} \\ \midrule
        \multicolumn{2}{c|}{CosyVoice2} & \textbf{4.10} $\pm$ \footnotesize{\textbf{1.09}} & 3.79 $\pm$ \footnotesize{1.36} & 4.17 $\pm$ \footnotesize{0.65} & 4.06 $\pm$ \footnotesize{0.85} & 3.31 $\pm$ \footnotesize{1.53} & 2.90 $\pm$ \footnotesize{1.31} \\
        \multicolumn{2}{c|}{F5TTS} & 3.97 $\pm$ \footnotesize{1.04} & 3.85 $\pm$ \footnotesize{1.14} & 2.94 $\pm$ \footnotesize{1.05} & 2.64 $\pm$ \footnotesize{0.97} & 2.94 $\pm$ \footnotesize{1.18} & 2.10 $\pm$ \footnotesize{1.06} \\ \midrule
        \multicolumn{2}{c}{DeepASMR} & 3.94 $\pm$ \footnotesize{1.14} & \textbf{4.13} $\pm$ \footnotesize{\textbf{1.11}} & \textbf{4.19} $\pm$ \footnotesize{\textbf{0.74}} & \textbf{4.17} $\pm$ \footnotesize{\textbf{0.74}} & \textbf{4.02} $\pm$ \footnotesize{\textbf{0.83}} & \textbf{3.97} $\pm$ \footnotesize{\textbf{0.96}} \\
       \midrule \midrule
        \multicolumn{8}{c}{\textit{Cross-style Synthesis}} \\ \midrule
        & & \multicolumn{2}{c|}{ASMR $\rightarrow$ Normal} & \multicolumn{4}{c}{Normal $\rightarrow$ ASMR} \\
        TTS Model & VC Model & \multicolumn{1}{c|}{ZH} & \multicolumn{1}{c|}{EN} & \multicolumn{2}{c|}{ZH} & \multicolumn{2}{c}{EN} \\
        &  & OI-MOS & OI-MOS & OI-MOS & ASMR-MOS & OI-MOS & ASMR-MOS  \\
        \midrule
        \multicolumn{2}{c|}{Ground Truth} & 4.04 $\pm$ \footnotesize{1.24} & 4.21 $\pm$ \footnotesize{1.12} & 4.35 $\pm$ \footnotesize{0.64} & 4.24 $\pm$ \footnotesize{0.66} & 4.36 $\pm$ \footnotesize{0.73} & 4.18 $\pm$ \footnotesize{0.80} \\ \midrule
                CosyVoice2 &  \multirow{2}{*}{CosyVoiceVC}  & 3.53 $\pm$ \footnotesize{1.12} & 3.74 $\pm$ \footnotesize{1.00} & 2.39 $\pm$ \footnotesize{1.13} & 2.27 $\pm$ \footnotesize{1.01} & 3.81 $\pm$ \footnotesize{1.11} & 2.34 $\pm$ \footnotesize{1.23} \\
        F5TTS & & 3.37 $\pm$ \footnotesize{1.17} & 3.84 $\pm$ \footnotesize{0.97} & 2.99 $\pm$ \footnotesize{1.32} & 2.34 $\pm$ \footnotesize{1.17} & 3.91 $\pm$ \footnotesize{1.12} & 2.30 $\pm$ \footnotesize{1.23} \\ \midrule
        CosyVoice2 & \multirow{2}{*}{SeedVC}& 2.67 $\pm$ \footnotesize{1.14} & 3.10 $\pm$ \footnotesize{1.04} & 2.18 $\pm$ \footnotesize{1.21} & 1.79 $\pm$ \footnotesize{1.16} & 3.84 $\pm$ \footnotesize{1.18} & 2.03 $\pm$ \footnotesize{1.39} \\
        F5TTS & & 2.50 $\pm$ \footnotesize{1.03} & 3.30 $\pm$ \footnotesize{1.10} & 2.58 $\pm$ \footnotesize{1.35} & 1.92 $\pm$ \footnotesize{1.26} & \textbf{4.08} $\pm$ \footnotesize{\textbf{1.08}} & 2.12 $\pm$ \footnotesize{1.41} \\
        \midrule
        \multicolumn{2}{c|}{DeepASMR} & \textbf{3.76} $\pm$ \footnotesize{\textbf{1.30}} & \textbf{3.88} $\pm$ \footnotesize{\textbf{1.18}} & \textbf{4.19} $\pm$ \footnotesize{\textbf{0.61}} & \textbf{3.99} $\pm$ \footnotesize{\textbf{0.87}} & 3.91 $\pm$ \footnotesize{0.93} & \textbf{3.66} $\pm$ \footnotesize{\textbf{0.99}} \\
        \bottomrule
    \end{tabular}%
    \caption{Subjective Evaluation Results for Different Models under Intra and Cross-Style Synthesis Scenarios}
    \label{tab:human}
\end{table*}

\subsection{Ablation Study}
To validate the contributions of our architectural components, we conducted ablation studies focusing on three key aspects: the task prompt selection strategy, the composition of the training dataset, and the iterative inference refinement mechanism.

\subsubsection{Effectiveness of Task Prompt Selection via Virtual Speaker Pool}
As discussed in Section \ref{sec:task_selector}, cross-style synthesis risks "timbre leakage," where the output voice drifts toward the style reference rather than the target speaker. We proposed an automated similarity-based retrieval strategy that retrieves a vocal neighbor from a candidate pool to mitigate this. Table ~\ref{tab:ablation_study_selector} compares four strategies: using a single high-quality female real utterance ("Real $\#1$"), using a high-quality fixed female real utterance and a fixed male utterance ("Real $\#2$"), retrieving from a pool of 50 real manually selected utterances ("Real $\#50$") including both male and female, and retrieving from our proposed pool of 50 synthetic utterances ("Virtual $\#50$").

The results validate the necessity of a retrieval-based approach. Using a single high quality female prompt ("Real  $\#1$") yields the lowest speaker similarity in the Normal-to-ASMR task, confirming that an arbitrary style reference degrades identity preservation. In contrast, expanding the search space to 50 candidates significantly improves identity retention. Crucially, our virtual pool achieves performance comparable to the real speaker pool, demonstrating that synthetic data can effectively serve as a style reference bank. This confirms that the virtual speaker pool offers a scalable, privacy-preserving alternative to mining real data without compromising synthesis fidelity.

\begin{table}[t]
    \centering
    \begin{tabular}{lcccccc}
        \toprule
        \multirow{2}{*}{Speaker Pool} & \multicolumn{3}{c}{ASMR $\rightarrow$ Normal} & \multicolumn{3}{c}{Normal $\rightarrow$ ASMR} \\
        & WER & SIM & Style & WER & SIM & Style \\
        \midrule
        Real \#1           & 1.34          & 0.24          & -0.64          & 12.27         & 0.27          & \textbf{+0.89} \\
        Real \#2      & 1.32          & 0.28          & -0.80          & 8.12          & \textbf{0.43} & +0.23 \\
        Real \#50    & \textbf{1.12} & 0.30          & \textbf{-0.88} & 11.19         & 0.42          & +0.45 \\
        Virtual \#50  & 1.90          & \textbf{0.31} & -0.82          & \textbf{6.53} & 0.41          & +0.58 \\
        \bottomrule
    \end{tabular}
\caption{Performance comparison for Cross-style Synthesis with different Speaker Pool}
\label{tab:ablation_study_selector}
\end{table}

\subsubsection{Impact of Training Dataset Composition}

We investigated the necessity of mixing normal speech data during the fine-tuning phase. Table \ref{tab:ablation_study_data} compares the pretrained model, a model fine-tuned exclusively on DeepASMR-DB against a model fine-tuned on a mixture of DeepASMR-DB and the Emilia (normal speech) dataset~\cite{he2024emilia}.

The results in Table \ref{tab:ablation_study_data} indicate that including normal speech is critical for model stability. We acknowledge that due to the limited number of speakers, fine-tuning exclusively on the ASMR dataset causes catastrophic forgetting of normal speech patterns, resulting in performance degradation on unseen speakers as evidenced by high WER and lower SIM scores. Integrating the Emilia dataset drastically reduces the N2N WER, restoring the model's general capability. Furthermore, this mixed strategy also benefits the Normal-to-ASMR task, reducing the WER from 15.20\% to 6.53\%. This suggests that maintaining a foundation of modal speech helps the model better interpret the linguistic content of normal prompts before converting them to the ASMR style.

\begin{table}[t]
\centering
\setlength{\tabcolsep}{2.2pt}
\begin{tabular}{c|c|cc|cc|cc|cc}
\toprule
\multirow{2}{*}{DeepASMR-DB} & \multirow{2}{*}{Emilia} &\multicolumn{2}{c|}{N2N} & \multicolumn{2}{c|}{A2A} & \multicolumn{2}{c|}{A2N} & \multicolumn{2}{c}{N2A} \\
& & WER & SIM & WER & SIM & WER & SIM & WER & SIM \\ \midrule
\multicolumn{2}{c|}{Pretrained Model} & 2.10 & 0.61 & 3.20 & 0.49 & 1.40 & 0.29 & 2.60 & 0.44 \\ \midrule
\checkmark & \ding{55} &3.30 & 0.53& 9.30 & 0.47 & 2.80 & 0.29 & 15.20 & 0.37 \\ 
\checkmark & \checkmark & \textbf{1.91} & \textbf{0.59} & \textbf{6.72} & \textbf{0.47} & \textbf{1.90} & \textbf{0.31} & \textbf{6.53} & \textbf{0.41} \\ \bottomrule
\end{tabular}
\caption{Performance Comparison with Different Training Dataset Composition in the Fine-tuning Stage}
\label{tab:ablation_study_data}
\end{table}

\subsubsection{Trade-offs in Iterative Inference Refinement}

For challenging cross-style scenarios, we proposed an iterative refinement strategy where the output of the first pass serves as the speaker prompt for subsequent passes. Table \ref{tab:ablation_study_repeat} quantifies the impact of repeating this process up to three times. 

We observe a distinct trade-off between style intensity and speaker identity. The first iteration (Step 1) provides a balanced baseline with a Style score of +0.58 and a SIM score of 0.41. Proceeding to Step 2 significantly enhances the ASMR texture, raising the Style score to +0.80 and improving intelligibility. However, this comes at the cost of speaker similarity, which drops to 0.29. A third iteration yields diminishing returns: while the Style score marginally increases, the WER degrades to 5.72\%, and similarity falls further to 0.25. 

Conversely, for the ASMR-to-Normal task, the performance metrics remain remarkably stable. We attribute this robustness to the strong acoustic nature of normal speech. Since normal speech is characterized by distinct vocal fold vibrations and formant structures—which are the primary carriers of speaker identity—the speaker information in the generated normal speech is inherently strong and stable. Once the model reconstructs these voiced features in the first step, subsequent iterations do not degrade the timbre, unlike the N2A task where identity features are being actively suppressed.

Based on these findings, for all the experiments, we utilize step 1 as the default configuration, as our task requires a balance of intelligibility and identity preservation. However, for scenarios where the unvoice speech and ASMR sensation is primary and speaker resemblance is secondary, utilizing larger iterative refinement steps serves as an effective method to generate highly intense, authentic ASMR content.

\begin{table}[t]
    \centering
   \begin{tabular}{c|ccc|ccc}
\toprule
\multirow{2}{*}{Iterative Steps} & \multicolumn{3}{c|}{ASMR $\to$ Normal} & \multicolumn{3}{c}{Normal $\to$ ASMR} \\
 & WER & SIM & Style & WER & SIM & Style \\
\midrule
1 & 1.90  & 0.31 & -0.82 & 6.53 & \textbf{0.41} & +0.58 \\
2 & 2.15 & 0.31 & \textbf{-0.96} & \textbf{4.15} & 0.29 & +0.80  \\
3 & \textbf{1.55} & 0.31 & -0.81 & 5.72 & 0.25 & \textbf{+0.88} \\
\bottomrule
\end{tabular}
\caption{Performance Comparison for Cross-style Synthesis with Different Iterative Refinement Steps}
\label{tab:ablation_study_repeat}
\end{table}

\subsection{Unvoiced Speech Analysis}
A primary acoustic hallmark of ASMR-style speech is the transition from voiced phonation to unvoiced whispering (the suppression of vocal fold vibration). In technical terms, this involves a shift from a periodic glottal source to an aperiodic, turbulent noise source, resulting in the absence of a distinct fundamental frequency ($F_0$)~\cite{cirillo2004communication}. During whispering, the vocal folds do not vibrate; instead, they remain open, replacing the harmonic structure of normal speech with broadband noise.

Table \ref{tab:unvoiced_ratio_analysis} presents a quantitative evaluation of the models' ability to synthesize unvoiced speech (N2A) and revert to voiced speech (A2N) by measuring the Global Unvoiced Ratio ($R_{UV}$). DeepASMR demonstrates superior fidelity to these baselines across both tasks. In the N2A task, it achieves an unvoiced ratio of 74.21\%, significantly outperforming the cascade VC baselines. This discrepancy indicates that the VC-based approaches struggle to suppress vocal fold vibration, often reverting to voiced speech rather than generating pure whisper. Conversely, in the A2N task, DeepASMR closely approximates the natural voicing distribution of normal speech. In contrast, the cascade models exhibit inflated unvoiced ratios for the A2N task, suggesting that they fail to fully recover voicing from whispered inputs, resulting in output that remains perceptibly breathy or noisy. 

\begin{table}[t]
    \setlength{\tabcolsep}{8.2pt}
\centering
\begin{tabular}{c|c|c|c}
\toprule
TTS Model & VC Model & A2N & N2A \\
\midrule
\multicolumn{2}{c|}{GroundTruth} & 33.80\% & 91.78\% \\
\midrule
CosyVoice2 & \multirow{2}{*}{CosyVoiceVC} & 57.23\% & 37.35\% \\
F5TTS & & 58.84\% & 21.41\% \\
\midrule
CosyVoice2 & \multirow{2}{*}{SeedVC} & 56.63\% & 14.02\% \\
F5TTS & & 65.93\% & 14.80\% \\
\midrule
\multicolumn{2}{c|}{DeepASMR} &  \textbf{39.99}\% & \textbf{74.21}\% \\
\bottomrule
\end{tabular}
\caption{Comparison of Global Unvoiced Ratio for Cross-Style Synthesis}
\label{tab:unvoiced_ratio_analysis}
\end{table}

\subsection{Comparison with Commercial Models}

\begin{table}[t]
\centering
\begin{tabular}{l|c}
\toprule
Model & Global Unvoiced Ratio \\
\midrule
Prompt                       & 38.69\% \\
\midrule
CosyVoice2                   & 25.96\% \\
F5TTS                        & 22.35\% \\
ElevenLabs v3 Alpha                   & 35.42\% \\
MiniMax speech-hd-02                       & 25.54\% \\
\midrule
DeepASMR (Intra-Style) & 24.75\% \\
DeepASMR (Cross-Style) & \textbf{73.76\%} \\
\bottomrule
\end{tabular}
\caption{Comparison of Global Unvoiced Ratio with Commercial Models}
\label{tab:unvoiced_ratios}
\end{table}
ElevenLabs v3 Alpha~\cite{elevenlabs} and MiniMax speech-hd-02~\cite{zhang2025minimax} are among the leading commercial large-scale TTS models. Their core strengths lie in excellent speech naturalness, strong emotional expressiveness, and extensive multilingual support. While their specific model architectures and training data remain undisclosed, both platforms support ASMR-style generation for fixed, predefined speakers.

For this comparison, we utilized a total of 8 different prompts sourced from their official API platforms. It is important to note that these reference prompts are dominated by voiced speech rather than unvoiced speech. Our analysis reveals that ASMR speech synthesized by these commercial models is predominantly based on vocal cord vibrations. However, this represents only a limited subset of the ASMR domain. A substantial portion of authentic ASMR involves unvoiced speech with minimal vocal cord vibration. Commercial models currently struggle to replicate this characteristic, particularly when dealing with unseen speakers.

Table~\ref{tab:unvoiced_ratios} presents the Global Unvoiced Ratio for each model, where a higher ratio indicates that the speech is more unvoiced. We evaluate our proposed method, DeepASMR, on two distinct tasks—intra-style synthesis and cross-style synthesis—and report results for both. As shown, DeepASMR (particularly in cross-style settings) achieves a significantly higher unvoiced ratio compared to commercial baselines. We invite you to listen to our demos to have a clearer comparison with ElevenLabs and MiniMax\footnote{https://vivian556123.github.io/deepasmr-demo/}.

\section{Conclusion and Future Work}
This paper introduces DeepASMR, the first framework to generate high-quality, personalized ASMR speech from any speaker's ordinary voice without prior ASMR samples. By leveraging token-level soft factorization, DeepASMR redefines style transfer; rather than forcing a rigid separation of content and style, our two-stage pipeline and task prompt selection explicitly manage residual timbre information. This transforms timbre from a source of leakage into a mechanism for consistent identity preservation. Supported by the 670-hour DeepASMR-DB, our experiments demonstrate that this controllable entanglement allows for the precise synthesis of unvoiced ASMR textures while retaining the sonic signature of the original speaker, surpassing existing baselines.

While our results are promising, it is important to note that ASMR is not a universal phenomenon, with emerging evidence pointing to subtle neuroanatomical distinctions between responsive and non-responsive individuals~\cite{fredborg2017examination}. Furthermore, our current stimulus design is limited to vocal speech. Future research will incorporate non-vocal triggers, such as rubbing or tapping, to broaden the scope of synthesis.

Finally, DeepASMR offers a versatile and scalable solution for media production and virtual agents. However, the ability to synthesize speech that maintains speaker identity carries potential risks, such as voice spoofing or impersonation. To mitigate these risks, future deployment should be accompanied by detection models designed to discriminate whether an audio clip was synthesized by DeepASMR.

\bibliographystyle{IEEEtran}
\bibliography{IEEEexample}


 

\vfill

\end{document}